\documentstyle[times,amsmath,graphicx]{jaa}

\begin{document}
\title[Cosmological N-body simulations: size matters]
{Effects of the size of cosmological N-body
                          simulations on physical quantities -II. \\ 
                          Halo formation and destruction rate} 
\author[J. Prasad]%
       {Jayanti Prasad\thanks{e-mail:jayanti@mri.ernet.in} \\ 
        Harish-Chandra Research Institute, Allahabad 211019, India}
\maketitle
\label{firstpage}
\begin{abstract}
In this study we show  how errors  due to finite box size affect formation 
and destruction rate for haloes in cosmological N-body simulations. In an 
earlier study we gave an analytic  prescription of finding the corrections 
in the mass function. Following the same approach, in this paper we give 
analytical expressions for corrections in the formation rate,  destruction 
rate  and the rate of change in comoving number density, and compute  their 
expected values for the power law ($n=-2$) and LCDM models.
\end{abstract}

\begin{keywords}
Large scale structure  of Universe -- N-Body simulations  -- galaxies
\end{keywords}

{\centering{\section{Introduction}}}
\label{sec:intro}
In prevalent models of structure formation in the universe, galaxies, clusters
of galaxies and other large scale structures are believed to have formed due 
to gravitational amplification of small perturbations (Peebles 1980; 
Padmanabhan 1993;  Peacock 1999; Padmanabhan 2002; Bernardeau {\it\ et al}. 
2002). In cold dark matter dominated models these perturbations collapse 
hierarchically i.e., smaller  structures form first and then they merge and 
form larger structures.

In the models of structure formation the main goal is to understand the growth
of perturbations at various scales. It has been found that the linear 
perturbation theory closely follows the actual growth of perturbation at any 
scale as long as the amplitude of perturbation at that scale is small. 
However, the linear approximation breaks down once the amplitude becomes large
because perturbations at various scales couple to each other and the system 
becomes nonlinear. Many approximations have been proposed (Zel'dovich 1970; 
Gurbatov {\it\ et al}. 1989;  Saichev $\&$ Shandarin 1989; Matarrese 
{\it et al}. 1992; Brainerd {\it et al}. 1993; Hui $\&$ Bertschinger 1996; 
Bagla $\&$ Padmanabhan 1994; Sahni $\&$ Coles 1995) to  explain  nonlinear  
gravitational clustering but their validity has been limited to  special 
cases. Cosmological N-body simulations are the main tools to understand the 
gravitational clustering in nonlinear regime  (Efstathiou G., {\it\ et al}. 
1985; Bertschinger 1998; Bagla $\&$ Padmanabhan 1997a; Bagla 2005). 

In N-body simulations we simulate a representative  region of the universe 
which is in general large but finite. This restricts us from incorporating 
perturbations at scales greater than the size of the simulation box.
We also  cannot consider perturbations  at scales smaller than the grid
length. This means that there is a natural truncation of power spectrum at 
large as well as small wave numbers. It has been shown in the earlier studies
 (Peebles 1974, 1985; Little, Weinberg $\&$ Park 1991; Bagla $\&$ Padmanabhan 
1997b) that the truncation of power spectrum at small scales does not affect 
large scale clustering in any significant way. However, significant effects 
of the  truncation of power spectrum at large scales on clustering at small 
scales have been  found (Gelb $\&$ Bertschinger 1994a,b; Bagla $\&$ Ray 
2005; Power $\&$ Knebe 2006; Bagla $\&$  Prasad  2006, hereafter BP06).

Apart from N-body simulations, an analytical prescription given by Press $\&$ 
Schechter (1974) has been  used to find  mass function and related quantities. 
In this prescription the fraction of mass in the collapsed objects, having 
mass greater than a certain value, is identified with the fraction of volume
in the initial density field which had density contrast, filtered over an 
appropriate scale, greater than a critical  density contrast. The 
Press-Schechter formalism has been used in many studies to compute 
merging and mass function (Bond {\it\ et al}. 1991; Bower 1991; Lacey $\&$ 
Cole 1993, 1994; Sasaki 1994; Kitayama $\&$ Suto 1996; Cohn, Bagla $\&$ White
2001). In many of these studies the main goal has  been to test the 
results based on the Press-Schechter formalism against N-body simulations.
Here our goal is to find the analytic corrections in the formation and 
destruction rate of haloes in N-body simulations due to finite box size of
the simulation box.

The effects of the box size on the physical quantities in N-body simulations 
have been studied before. Gelb $\&$ Bertschinger (1994a) showed that 
the rms fluctuation in the mass and pairwise velocity dispersion at a given 
scale are underestimated  when the size of the simulation box is reduced. 
Bagla $\&$ Ray (2005) computed the finite box effects on the average two point
correlation function, power spectrum and cumulative mass function in N-body 
simulations. They found that the required box size for simulating the LCDM 
model at high redshift is much  larger  than what people generally use. Power 
$\&$ Knebe (2006) studied the effects of box size on kinematical properties of
cold dark matter haloes and concluded that the distribution of internal 
properties of  haloes (like spin parameter) is  not very sensitive to the box
size. So far most of the studies used  N-body simulations for drawing  their 
conclusions. In one of our earlier studies (BP06) we  proposed an analytic 
prescription for estimating the corrections due to box  size in N-body 
simulations for  rms fluctuations in mass, two  point correlation  function 
and the mass function. 

The mass function is an important physical quantity in nonlinear 
gravitational clustering. It  has been used to compare  various theoretical 
models with observations (Bagla, Padmanabhan \& Narlikar 1996; White, 
Efstathiou \&  Frenk 1993; Ostriker \& Steinhardt  1995; Ma \& 
 Bertschinger 1994; White 2002). However, it does not contain all the 
information  which is needed  for various physical processes. For example, if 
we want to know the comoving number density of ionizing sources at any 
redshift then the  mass function contains insufficient information  since not 
all bound objects become the source of ionization, only those objects become 
ionizing sources which form during a particular period of time (Chiu \& 
Ostriker 2000). In this situation we need to know the formation and 
destruction rate of collapsed objects as a  function of redshift. The rate 
with which quasars form in the early universe depend in an important way on 
the merger rate which is  another manifestation  of the formation rate 
(Carlberg 1990). Apart from these there are many other cases like 
understanding the epoch of cluster formation etc., for which formation and 
destruction rate of haloes are important (Bower 1991). In the present study we
 carry forward our program and  give analytic expressions  for the corrections
 in the formation and destruction rate  and estimate their values for the 
power law  ($n=-2$) and LCDM models.

The plan of this paper is as follows. In section 2 we give the basic equations 
which are needed for our analysis. We give the expression for the clustering
amplitude (rms fluctuations in mass),  mass function and the rate of change in
the comoving number density due to formation and destruction of haloes. In 
section 2.3 we show that the rate of change of comoving number density around 
a mass $M$ can be written in terms of the rate at which the haloes of that 
mass  are formed and the rate at which the haloes of that mass are destroyed 
due to  merging. We present our results in section 3 and discuss their 
implications for the power law and LCDM models. In section 4 we summarize our
 results.

\vspace{1.0cm}

{\centering{\section{Basic Equations}}}

{\centering{\subsection{Clustering amplitude}}}

In order to estimate the effects of box size on physical quantities in the 
linear regime, we use the mass variance  $\sigma^2(r)$ (Peebles 1993;
Peacock 1998; Padmanabhan 2002) as the base quantity.

\begin{equation}
\sigma^2(r) = 9\int \frac {k^3P(k)}{2\pi^2}
 \left ( \frac {\sin kr -kr\cos kr}{k^3r^3} \right)^2\frac{dk}{k}  
\label{eqn1}
\end{equation}
 On the basis of the correction in  $\sigma^2(r)$, we find the corrections in 
other physical quantities. In BP06, we showed that the variance in mass is
suppressed at all scales when we reduce the size of the simulation box. For 
example, if its actual value at scale $r$ is $\sigma_0^2(r)$ then in the 
initial conditions of a simulation   we obtain $\sigma^2(r,L_{Box})$ where: 

\begin{equation}
\sigma^2(r,L_{Box}) = \sigma^2_0(r) - \sigma_1^2(r,L_{Box})
\label{eqn2}
\end{equation} 

Here $L_{Box}$ is the size of the simulation box and $\sigma_1^2(r,L_{Box})$ 
is the correction  due to the finite box size. Note that here
$\sigma^2_1(r,L_{Box})$  is a positive quantity so clustering amplitude
$\sigma^2(r)$ is always underestimated when we reduce the size of the
simulation box.

\vspace{0.5cm}

{\centering{\subsection{Mass function and number density }}}

In the Press-Schechter formalism,  we consider the  initial density field 
as Gaussian random and smooth it over a filter of size $r$ (or mass $M$), then 
the fraction of mass in the collapsed objects $F(>M)$, having mass greater 
than $M$, at the final epoch, can  be identified with the fraction of volume 
in the initial density field which had  smoothed density contrast greater than
some critical density contrast $\delta_c$ which is computed on the basis of 
the spherical collapse model.
 
\begin{equation}
F(>M,t)  =  \frac{2}{\sqrt{\pi}} 
       \int\limits_{\frac{\delta_c (t)}{{\sqrt 2}\sigma(M)}}^{\infty}e^{-x^2}
 dx         =  erfc\left ( \frac{\delta_c (t)}{\sqrt{2}\sigma(M)} \right )
\label{eqn3}
\end{equation}

The comoving number density $N_{PS}(M,t)dM$ of objects which have mass in
the range $[M,M+dM]$ at time $t$ is given by

\begin{align}
N_{PS}(M,t)dM
& = \frac {\rho_0}{M} \times  \frac {dF(>M)}{dM} dM \nonumber \\
& = \sqrt{ \frac{2}{\pi}} \frac{\rho_0}{M} 
\left(-\frac{\delta_c(t)}{\sigma^2(M)} \frac{d\sigma(M)}{dM}\right )
 \exp \left (-\frac{\delta_c^2(t)}{2\sigma^2(M)}\right)dM
\label{eqn4}
\end{align}

Here we can use $\delta_c(t)= \delta_c/D(t)$ where $D(t)$ is the linear growth
factor which depends on the cosmological model being considered and 
$\delta_c(t)$ is taken to be 1.68 at the present epoch. 

\vspace{0.5cm}
{\centering{\subsection{The rate of change of number density}}}

We can find the rate of  change in the comoving number number density per unit 
time  for objects which have mass in the range [$M,M+dM$] from 
equation~(\ref{eqn4}).

\begin{align}
\left(\frac {dN_{PS}(M,t)}{dt} \right)dM
&= \sqrt{ \frac{2}{\pi}} \frac{\rho_0}{M} 
\left (\frac{1}{D^2(t)}\frac{dD(t)}{dt} \right)
\left (\frac{\delta_c}{\sigma^2(M)} \frac{d\sigma(M)}{dt} \right) \nonumber \\ 
& \times   \left [ 1- \frac {\delta_c^2}{\sigma^2(M)D^2(t)} \right ]
\exp \left (- \frac{\delta_c^2}{2\sigma^2(M)D^2(t)} \right )dM 
\label{eqn5}
\end{align}

We can identify the first and second terms of the right hand side with 
the destruction and formation rate  respectively  (Sasaki 1994; Kitayama 
$\&$ Suto 1996).

\begin{align}
\left(\frac{dN_{PS}(M,t)}{dt}\right) dM & = - \frac {1}{D(t)}\frac {dD(t)}{dt} 
 \left [ 1- \frac {\delta_c^2}{\sigma^2(M)D^2(t)} \right ]N_{PS}(M,t)dM
 \nonumber \\ 
& = - \left( \frac{d N_{Dest}(M,t)}{dt}\right) dM + 
 \left( \frac {dN_{Form}(M,t)}{dt} \right) dM
\label{eqn5a}
\end{align}

The formation rate $(dN_{Form}(M,t)/dt)dM$ quantifies the change in the 
comoving number density of objects around mass $M$, per unit time, due to the
formation of objects in that mass range when objects of mass smaller 
than $M$ merge together.

\begin{align}
\left(\frac{dN_{Form}(M,t)}{dt}\right) dM
&  =  \frac {1}{D(t)}\frac {dD(t)}{dt} 
  \left [\frac {\delta_c^2}{\sigma^2(M)D^2(t)} \right ]N_{PS}(M,t)dM
\nonumber \\
&= \sqrt{ \frac{2}{\pi}} \frac{\rho_0}{M} 
\left (\frac{1}{D^4(t)}\frac{dD(t)}{dt} \right)
\left (-\frac{\delta_c^3}{\sigma^{4}(M)} \frac{d\sigma(M)}{dt} \right)
\nonumber \\ 
& \times \exp \left (- \frac{\delta_c^2}{2\sigma^2(M)D^2(t)} \right )dM
\label{eqn6}
\end{align}  

The destruction rate $(dN_{Dest}(M,t)/dt)dM$ quantifies the rate of change of
comoving number density of haloes in the mass range [$M,M+dM$] when  
the haloes in that mass range merge together and form bigger haloes.

\begin{align}
\left( \frac{dN_{Dest}(M,t)}{dt}\right) dM
&  = \frac {1}{D(t)}\frac {dD(t)}{dt} N_{PS}(M,t)dM
\nonumber \\
&= \sqrt{ \frac{2}{\pi}} \frac{\rho_0}{M} 
\left (\frac{1}{D^2(t)}\frac{dD(t)}{dt} \right)
\left (-\frac{\delta_c}{\sigma^2(M)} \frac{d\sigma(M)}{dt} \right)dM
\nonumber \\ 
& \times \exp \left (- \frac{\delta_c^2}{2\sigma^2(M)D^2(t)} \right )dM
\label{eqn7}
\end{align}  

\vspace{1.0cm}
{\centering{\section{Box corrections}}}

On the basis of the correction due to finite box size in $\sigma^2(r)$
(equation~(\ref{eqn2})) we can find the corrections in the number density,
the rate of change in the number density, the rate of formation and the rate 
of destruction  i.e., equation~(\ref{eqn4}), (\ref{eqn5}),  (\ref{eqn6}), and 
(\ref{eqn7}) respectively.

\vspace{0.5cm}
{\centering{\subsection{Corrections in comoving number density}}}

On the basis of the Press-Schechter formalism the comoving number density of
objects in a mass range around a mass scale can be related to the rms 
fluctuations in the mass at that scale. Using the technique which we have 
developed to compute the  corrections due finite box size in the rms 
fluctuations in mass (see equation~(\ref{eqn2})), we can find the corrections 
in the comoving number density also.

The comoving number density which we expect in cosmological N-body simulations
is given by

\begin{align}
N_{PS}(M,t)dM & = \sqrt{ \frac{2}{\pi}} \frac{\rho_0}{M} \frac{\delta_c}{D(t)}
\left(-\frac{1}{\sigma^2(M)} \frac{d\sigma(M)}{dM}\right ) \exp \left (-
\frac{\delta_c^2}{2\sigma^2(M)D^2(t)} \right )dM \nonumber \\
& = N_{{PS},0}(M,t)dM - N_{{PS},1}(M,t)dM
\label{eqn8}
\end{align}

Here $N_{{PS},0}(M,t)$ and $N_{PS}(M,t)$ are the theoretical (box size
infinite) and the actual (box size finite) comoving number densities 
respectively.

\begin{align}
N_{{PS},0}(M,t)dM 
&  = \sqrt{ \frac{2}{\pi}} \frac{\rho_0}{M} \frac{\delta_c}{D(t)}
\left(\frac{-1}{\sigma^2_0(M)} \frac{d\sigma_0(M)}{dM}\right ) \exp \left (
\frac{-\delta_c^2}{2\sigma^2_0(M)D^2(t)} \right )dM
\label{eqn9}
\end{align}

and

\begin{eqnarray}
N_{{PS},1}(M,t)dM
 = \frac{1}{2} \left [1- \left (1-\frac{\sigma_1^2}{\sigma_0^2}\right)^{-3/2}
\left(1-\frac{d\sigma_1^2}{d\sigma_0^2} \right ) \right.  \nonumber \\ \times 
\left.  \exp \left \{ -\frac{\delta_c^2}{2D^2(t)} \left( \frac{1}{\sigma^2}-
\frac {1}{\sigma_0^2} \right ) \right \} \right ]N_{{PS},0}(M,t)dM
\label{eqn10}
\end{eqnarray}

or

\begin{equation}
N_{{PS},1}(M,t) dM \approx 
\frac{1}{2}\left [ \frac{d\sigma_1^2}{d\sigma_0^2}- \frac{3\sigma_1^2}
{2\sigma_0^2} + \frac {\delta_c^2 \sigma_1^2}{2D^2(t)\sigma_0^4}
 \right ]N_{{PS},0}(M,t)dM~~\mbox{if}~~\sigma^2_1/\sigma_0^2 < 1
\label{eqn10a}
\end{equation}

In equation~(\ref{eqn10a}) the coefficient of $N_{{PS},0}(M,t)$ changes sign 
at the scale for which 
$\sigma_0 \approx \delta_c/D(t)\sqrt{3}=\delta_c(t)/\sqrt{3} $ 
so the number density of objects below this scale is overestimated and the 
number density above  this is scale underestimated. This is in accordance with
our earlier results (BP06).

\begin{figure}
\begin{center}
\includegraphics[width=3in]{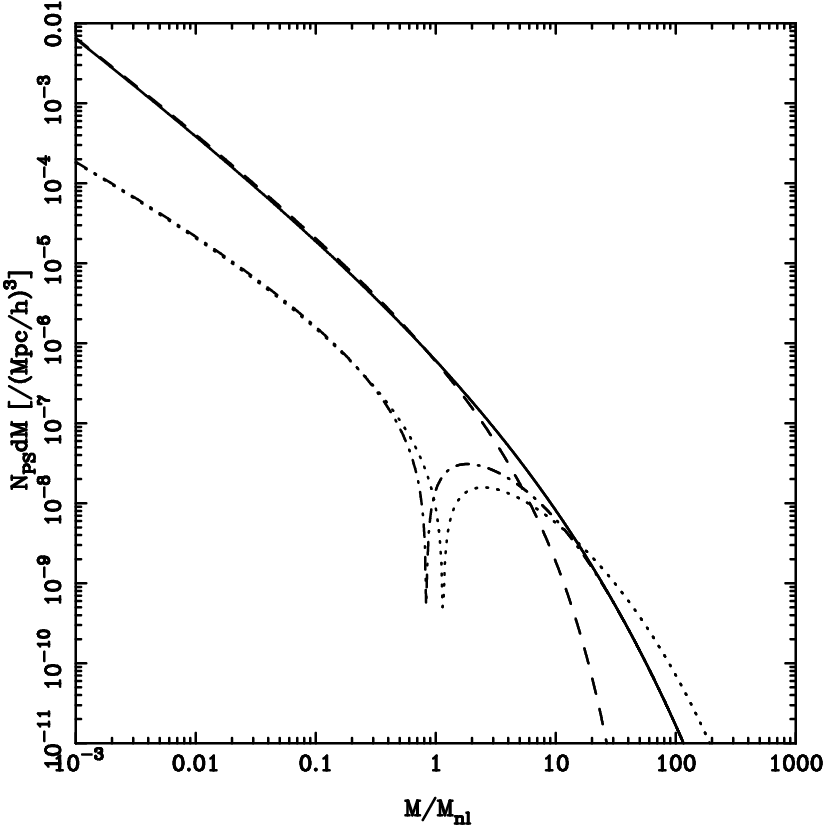}
\caption{This figure shows  the actual comoving number density $N_{PS,0}dM$ 
(solid line) and the comoving number density $N_{PS}~dM$ (dashed line) which 
we expect in a cosmological N-body simulation at the present epoch ($z=0$) 
for a power law ($n=-2$) model (see equation~(\ref{eqn8}, \ref{eqn9})).
Here we consider the size of the simulation  box $128~h^{-1}Mpc$ and normalize
the initial power spectrum such that the scale of non-linearity ($r_{nl}$) at  
$z=0$ is $8~h^{-1}Mpc$. In the figure, the exact (dot-dashed line) and approximate
  (dotted line) corrections in the comoving number density due to finite box size 
are also shown (see equation~(\ref{eqn10}, \ref{eqn10a})).}  
\end{center}
\label{fig1}
\end{figure}

\begin{figure}
\begin{center}
\includegraphics[width=3in]{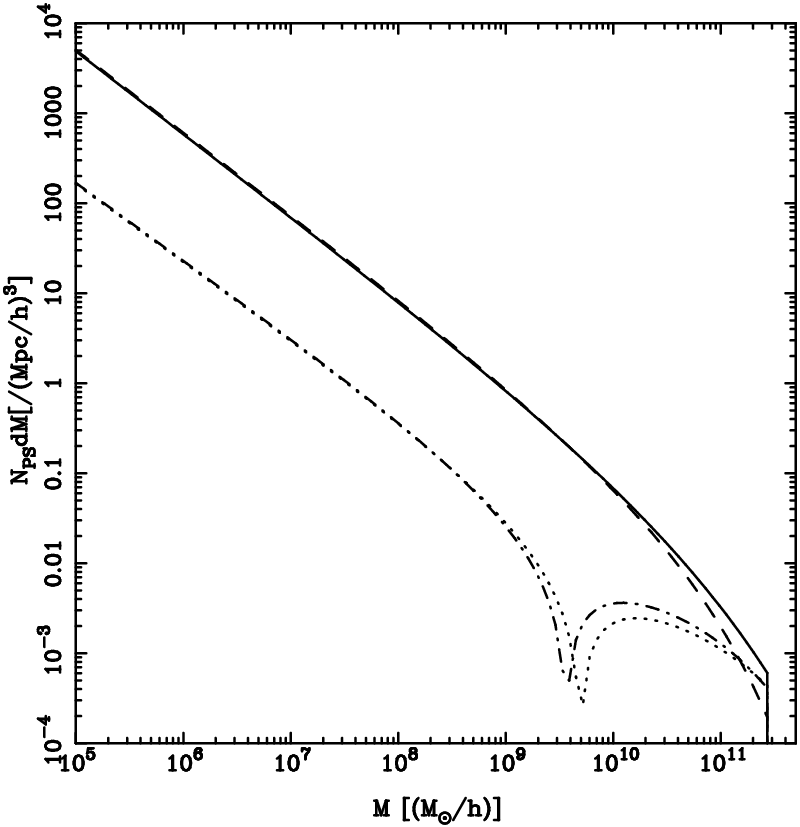}
\caption{This figure shows  the actual comoving number density $N_{PS,0}~dM$ 
(solid line) and the comoving number density $N_{PS}~dM$ (dashed line) which 
we expect in a cosmological N-body simulation for the  LCDM  model at $z=6$
(see equation~(\ref{eqn8}, \ref{eqn9})).
Here the size of the simulation is taken to be  $10~h^{-1}~Mpc$. We have also 
shown the exact (dot-dashed line) and  approximate (dotted line) corrections 
in the comoving number density due to finite box size 
(see equation~(\ref{eqn10}, \ref{eqn10a})) .}
\end{center}
\label{fig2}
\end{figure}

Figure (1) shows the theoretical comoving number density $N_{PS,0}$ and the 
actual comoving number density $N_{PS}$, which we expect in a N-body simulation
for the  power law model ($n=-2$) at the present epoch ($z=0$). In this case we consider 
a simulation box of size $128~h^{-1}Mpc$   and normalize the initial power 
spectrum such that the scale of non-linearity at $z=0$ is $8~h^{-1}Mpc$. 
The actual and approximate corrections in the comoving number density are also shown 
in the figure. This figures shows that in N-body simulations the  number density of 
large mass haloes is underestimated.  However, it is overestimated for small mass 
haloes. This feature is more evident from  the actual and approximate corrections 
(difference between the theoretical and actual values). 
In Figure (2) we show   the theoretical and the actual comoving number densities and  
corrections due to finite box size for the LCDM model. In this case we consider 
a  simulation box of size $10h^{-1} Mpc$ and compute physical quantities at $z=6$. 
From Figure (1) and Figure (2) it is clear that the overall trend  remains the same. 
This result is in  agreement with the  results of our earlier study (BP06).

\vspace{0.5cm}

{\centering{\subsection{Formation rate }}}

The rate of formation of haloes of mass $M$  which we expect in N-body 
simulation is given by  equation~(\ref{eqn6})

\begin{eqnarray}
\left( \frac{dN_{Form}(M,t)}{dt}\right) dM =  
\sqrt{ \frac{2}{\pi}} \frac{\rho_0}{M} 
\left (\frac{1}{D^4(t)}\frac{dD(t)}{dt} \right)
\left (-\frac{\delta_c^3}{\sigma^4(M)} \frac{d\sigma(M)}{dt} \right)
\nonumber \\ 
\times \exp \left (- \frac{\delta_c^2}{2\sigma^2(M)D^2(t)} \right )dM
\nonumber \\
= \left( \frac{dN_{Form,0}(M,t)}{dt}\right) dM- \left( \frac{dN_{Form,1}(M,t)}{dt}\right) dM
\label{eqn11}
\end{eqnarray}  

Here $(dN_{Form,0}(M,t)/dt)dM$  and $(dN_{Form,1}(M,t)/dt)dM$   are the 
theoretical formation rate and the correction term respectively. 

\begin{eqnarray}
\left( \frac{dN_{Form,0}(M,t)}{dt} \right) dM =  
\sqrt{ \frac{2}{\pi}} \frac{\rho_0}{M} 
\left (\frac{1}{D^4(t)}\frac{dD(t)}{dt} \right)dM  \nonumber \\ 
\times \left (-\frac{\delta_c^3}{\sigma^{4}_0(M)} \frac{d\sigma_0(M)}{dt} \right)
\exp \left (- \frac{\delta_c^2}{2\sigma^2_0(M)D^2(t)} \right )
\label{eqn12}
\end{eqnarray}  

and

\begin{eqnarray}
\left( \frac{dN_{Form,1}(M,t)}{dt}\right) dM
 = \frac{1}{2} \left [1- \left (1-\frac{\sigma_1^2}{\sigma_0^2}\right)^{-5/2}
\left(1-\frac{d\sigma_1^2}{d\sigma_0^2} \right ) \right.  \nonumber \\ \times
\left.  \exp \left \{ -\frac{\delta_c^2}{2D^2(t)} \left( \frac{1}{\sigma^2}-
\frac {1}{\sigma_0^2} \right ) \right \} \right ] \left( \frac{dN_{Form,0}(M,t)}{dt} \right) dM
\label{eqn13}
\end{eqnarray}

or

\begin{equation}
\left(\frac{dN_{Form,1}(M,t)}{dt}\right) dM\approx 
\frac{1}{2}\left [ \frac{d\sigma_1^2}{d\sigma_0^2}- \frac{5\sigma_1^2}
{2\sigma_0^2} + \frac {\delta_c^2 \sigma_1^2}{2D^2(t)\sigma_0^4}
 \right ] \left( \frac{dN_{Form,0}(M,t)}{dt} \right) dM
\label{eqn13a}
\end{equation}
$~~\mbox{if}~~\sigma^2_1/\sigma_0^2 < 1$ \\

\begin{figure}
\begin{center}
\includegraphics[width=3in]{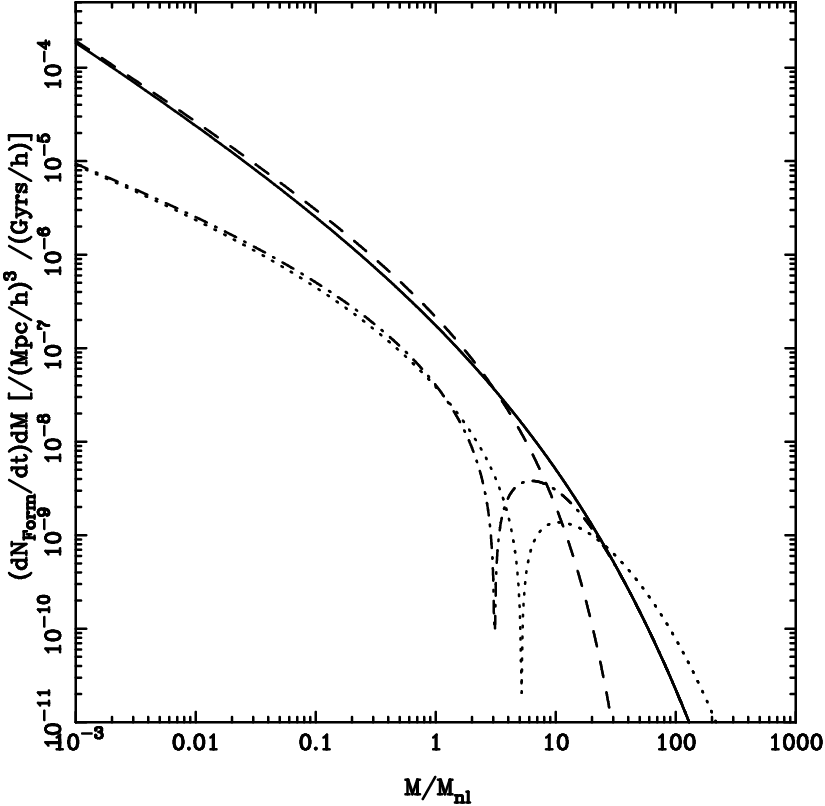}
\caption{This figure shows the actual formation rate $(dN_{Form,0}/dt) dM$ 
(solid line) and the formation rate  $(dN_{Form}/dt)~dM$ (dashed line) which 
we expect in a cosmological N-body simulation for the  power law  model 
(see equation~(\ref{eqn11}, \ref{eqn12})). All the parameters for this figure 
are the same as for  Figure 1. We have also shown the exact (dot-dashed line)
and  approximate  (dotted line) corrections  in the formation rate due to finite 
box  size (see equation~(\ref{eqn13}, \ref{eqn13a})).}
\end{center}
\label{fig3}
\end{figure}

\begin{figure}
\begin{center}
\includegraphics[width=3in]{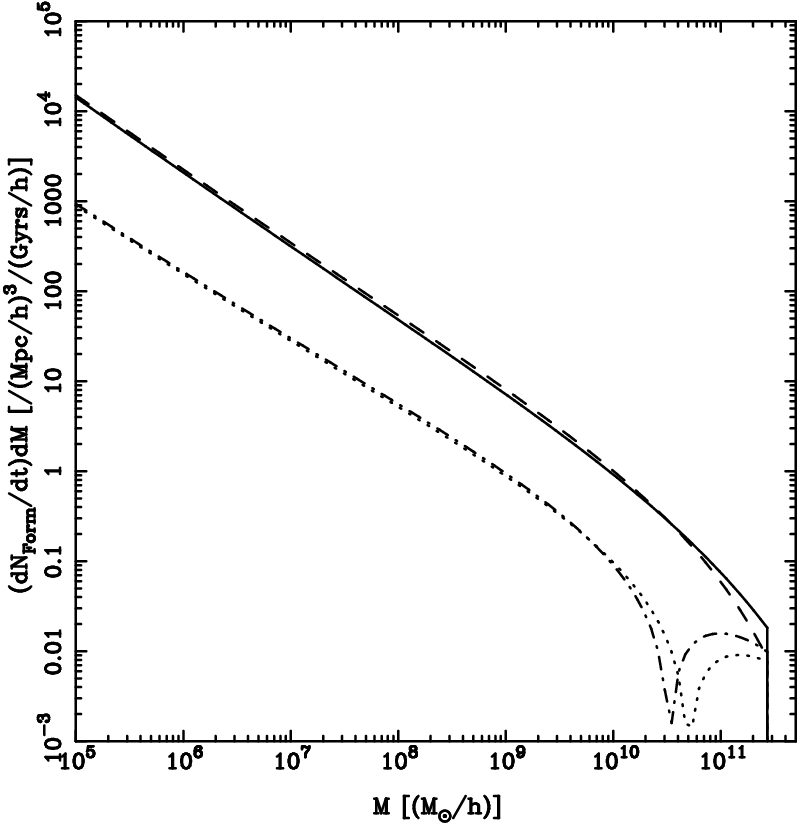}
\caption{This figure shows the actual formation rate  $(dN_{Form,0}/dt) dM$ 
(solid line) and the formation rate  $(dN_{Form}/dt)~dM$ (dashed line) which 
we expect in a cosmological N-body simulation for the LCDM  model 
(see equation~(\ref{eqn11}, \ref{eqn12})).  For this figure all the parameters are
identical to that of Figure 2. The  exact (dot-dashed line) and approximate 
(dotted line) corrections  in the formation rate  due to finite box size are also shown
(see equation~(\ref{eqn13}, \ref{eqn13a})).}
\end{center}
\label{fig4}
\end{figure}

Figure (3) and Figure (4) show the theoretical formation rate $(dN_{Form,0}/dt)dM$ 
and the actual formation rate $(dN_{Form}/dt)dM$ for the power law  and LCDM models
 respectively. The parameters for the power law and LCDM model are the same as in 
Figure (1) and Figure (2) respectively. Since the formation  rate is directly proportional 
to the comoving number density, it follows the same trend as the comoving number density.
 However, in this case the scale at  which the correction term changes sign is different 
from the scale at which the correction term for the comoving number density changes i.e., 
here it is the scale for which  $\sigma_0=\delta_c/D(t)\sqrt{5} =\delta_c(t)/\sqrt{5}$. 
From these figures we see that the formation rate of massive haloes is 
suppressed. However, that of low mass haloes is enhanced in N-body 
simulations when the box size is reduced. The main reason behind the 
suppression of the formation of large mass haloes is the absence of 
fluctuations in the initial density field at large scales due to truncation of
power.

\vspace{0.5cm}

{\centering{\subsection{Destruction rate }}}

Following the same approach as we have applied for the formation rate, we 
can find the corrections due to finite box size for destruction rate i.e., 
equation~(\ref{eqn7}) also. 

\begin{equation}
\left(\frac{dN_{Dest}(M,t)}{dt}\right) dM= \left( \frac{dN_{Dest,0}(M,t)}{dt}\right) dM
-\left( \frac{dN_{Dest,1}(M,t)}{dt}\right) dM
\label{eqn14}
\end{equation}

where 

\begin{eqnarray}
\left(\frac{dN_{Dest,0}(M,t)}{dt}\right)  dM=  
\sqrt{ \frac{2}{\pi}} \frac{\rho_0}{M} 
\left (\frac{1}{D^2(t)}\frac{dD(t)}{dt} \right) \nonumber \\ 
\times \left (-\frac{\delta_c}{\sigma^2_0(M)} \frac{d\sigma_0(M)}{dt} \right)dM
\exp \left (- \frac{\delta_c^2}{2\sigma^2_0(M)D^2(t)} \right )
\label{eqn15}
\end{eqnarray}  

and

\begin{eqnarray}
\left(\frac{dN_{Dest,1}(M,t)}{dt}\right) dM
 = \frac{1}{2} \left [1- \left (1-\frac{\sigma_1^2}{\sigma_0^2}\right)^{-3/2}
\left(1-\frac{d\sigma_1^2}{d\sigma_0^2} \right ) \right.  \nonumber \\ \times
\left.  \exp \left \{ -\frac{\delta_c^2}{2D^2(t)} \left( \frac{1}{\sigma^2}-
\frac {1}{\sigma_0^2} \right ) \right \} \right ] \left(\frac{dN_{Dest,0}(M,t)}{dt}\right)
\label{eqn16}
\end{eqnarray}

or

\begin{equation}
\left(\frac{dN_{Dest,1}(M,t)}{dt}\right) dM\approx 
\frac{1}{2}\left [ \frac{d\sigma_1^2}{d\sigma_0^2}- \frac{3\sigma_1^2}
{2\sigma_0^2} + \frac {\delta_c^2 \sigma_1^2}{2D^2(t)\sigma_0^4}
 \right ]  \left(\frac{dN_{Dest,0}(M,t)}{dt}\right)
\label{eqn16a}
\end{equation}
$~~\mbox{if}~~\sigma^2_1/\sigma_0^2 < 1$ \\

\begin{figure}
\begin{center}
\includegraphics[width=3in]{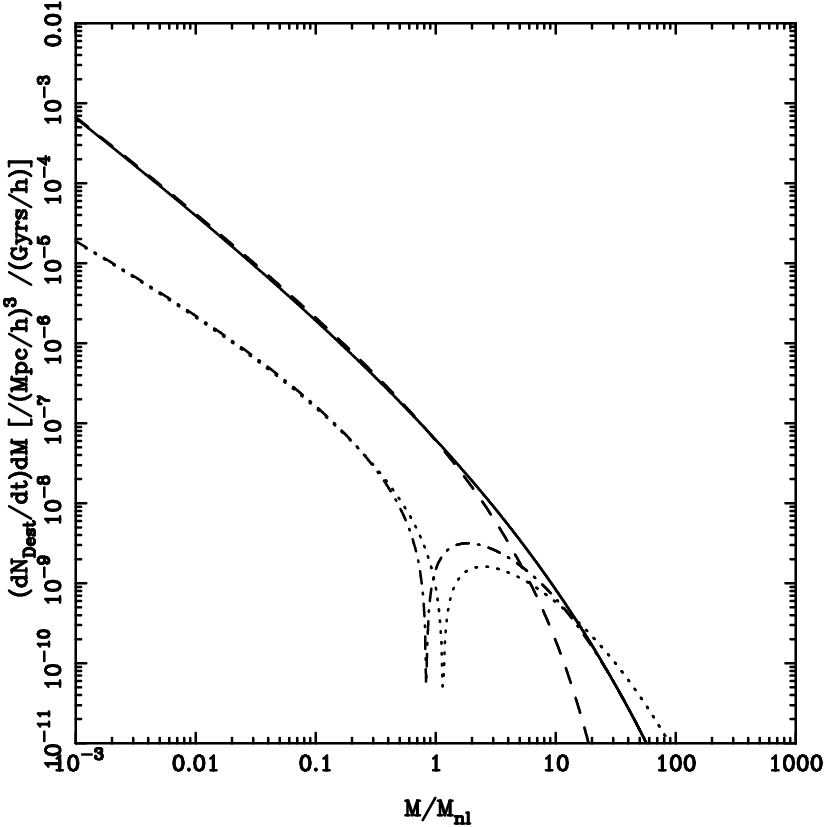}
\caption{This figure shows the actual destruction rate $(dN_{Dest,0}/dt) dM$ 
(solid line) and the formation rate  $(dN_{Dest}/dt)~dM$ (dashed line) which 
we expect in a cosmological N-body simulation (see equation~(\ref{eqn14}, 
\ref{eqn15})) for the power law model. All the parameters for this figure
are identical to that of Figure 1. We have also shown the exact (dot-dashed line) 
and  approximate  (dotted line) corrections in the destruction  rates  
due to finite box size (see equation~(\ref{eqn16},\ref{eqn16a})).}
\end{center}
\label{fig5}
\end{figure}

\begin{figure}
\begin{center}
\includegraphics[width=3in]{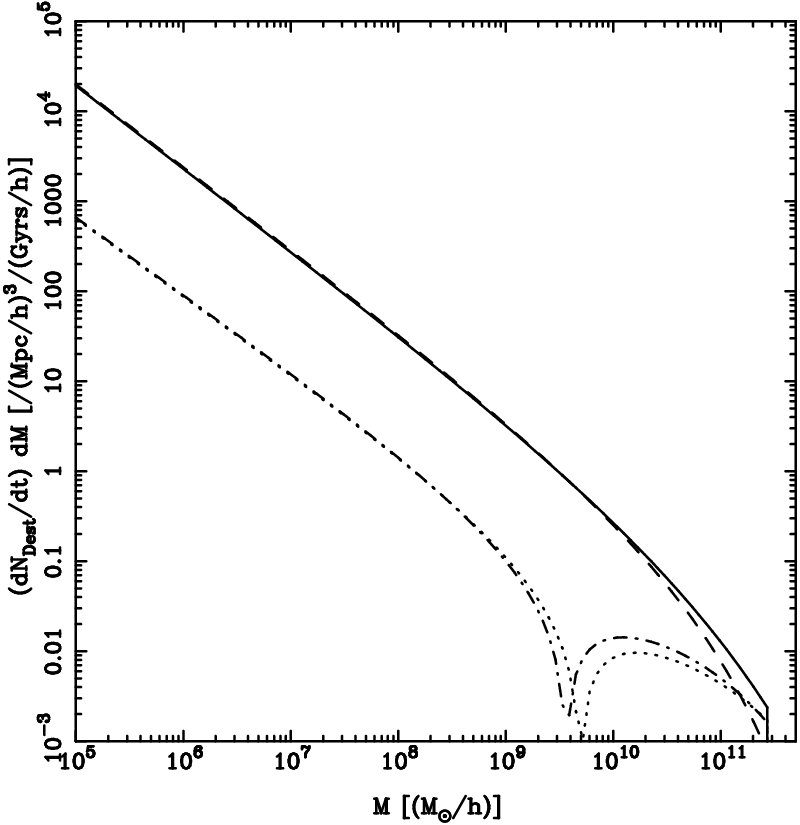}
\caption{This figure shows the actual destruction  rate  $(dN_{Dest,0}/dt) dM$ 
(solid line) and the formation rate  $(dN_{Dest}/dt)~dM$ (dashed line) which 
we expect in cosmological a N-body simulations(see equation~(\ref{eqn14}, 
\ref{eqn15}) for  the LCDM  simulation. All the parameters for this figure
are identical to that of Figure 2. The exact (dot-dashed line) and  
approximate  (dotted line) corrections in the destruction rate  due to finite 
box size are also plotted  (see equation~(\ref{eqn16},\ref{eqn16a})).}
\end{center}
\label{fig6}
\end{figure}

\begin{figure}
\begin{center}
\includegraphics[width=3in]{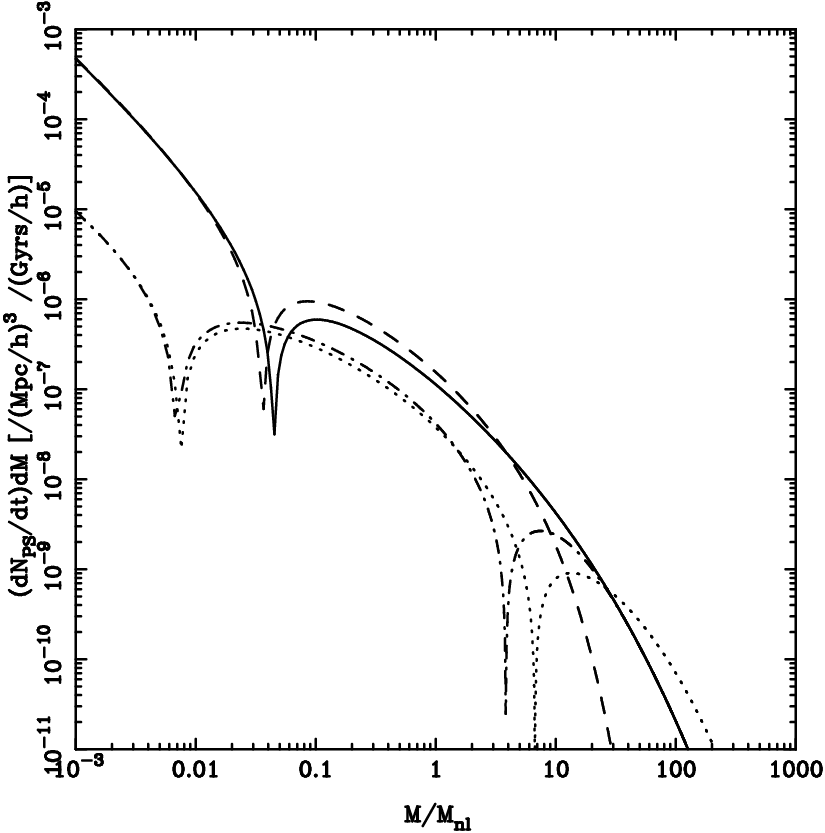}
\caption{This figure shows  the actual merger  rate $(dN_{PS,0}/dt) dM$ 
(solid line) and the formation rate  $(dN_{PS}/dt)~dM$ (dashed line) which we 
expect in a cosmological N-body simulation for the power law ($n=-2$)
(see equation~(\ref{eqn17})). All the parameters for this figure 
are the same as for  Figure 1. The exact (dot-dashed line) and  approximate 
 (dotted line) corrections  in the  rate of change in the comoving number density 
are also  plotted.} 
\end{center}
\label{fig7}
\end{figure}

\begin{figure}
\begin{center}
\includegraphics[width=3in]{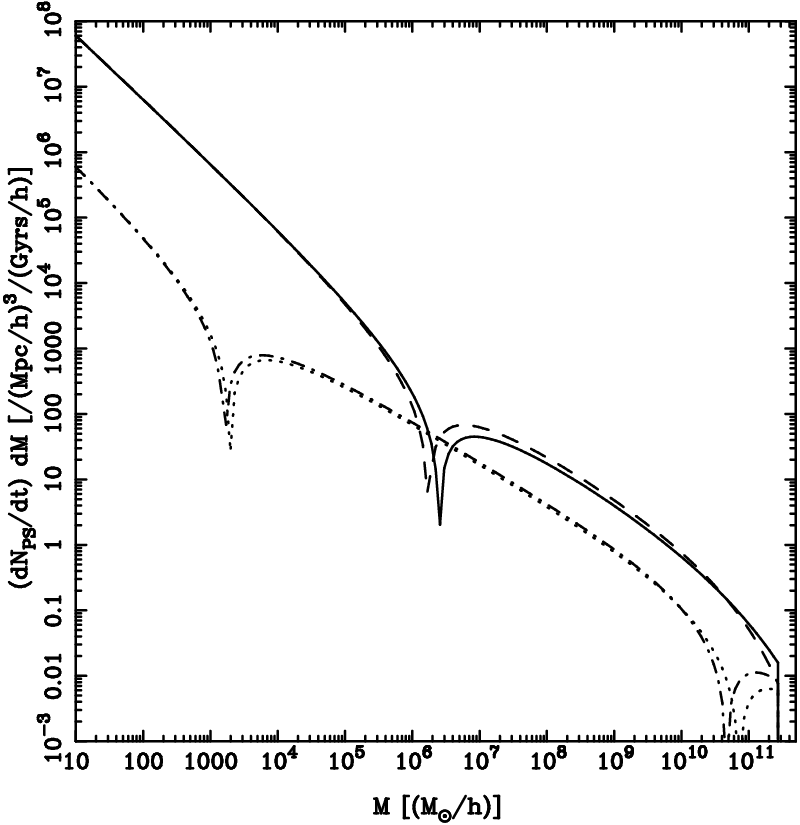}
\caption{This figure shows  the actual merger rate  $(dN_{PS,0}/dt) dM$ 
(solid line) and the formation rate  $(dN_{PS}/dt)~dM$ (dashed line) which we 
expect in a cosmological N-body simulation for the LCDM  model. 
 All the parameters for this figure are the same as for  Figure 2. The exact 
(dot-dashed line) and approximate (dotted line) corrections in the  rate of 
change in the comoving number density  are also  plotted.} 
\end{center}
\label{fig8}
\end{figure}

Figures (5) and (6) show the destruction rates for the power law and LCDM model
respectively. The parameters for the power law and LCDM models are the same as 
in  Figure (1) and Figure (2) respectively. In this case the correction 
term changes sign at the same scale at which the correction term for the 
comoving  number density changes. This is because $N_{PS,1}/N_{PS,0}$ and 
$N_{Dest,1}/N_{Dest,0}$ are equal and so $N_{PS,1}$ and $N_{Dest,1}$ have the 
same scale of zero crossing. Here also we find that the destruction rate  for 
the massive haloes are suppressed. However, they are enhanced for the low mass 
haloes when we reduce the size of the simulation box. 


\vspace{0.5cm}

{\centering{\subsection{Rate of change on number density}}}

The  rate of change of the  number density is defined as (see 
equation~(\ref{eqn6}) 

\begin{align}
\left( \frac{ N_{PS}(M,t)}{dt}\right) dM 
&=  \frac {-1}{D(t)}\frac {dD(t)}{dt} 
  \left [ 1- \frac {\delta_c^2}{\sigma^2(M)D^2(t)} \right ]N_{PS}(M,t)dM
\nonumber \\ 
& = - \left( \frac{dN_{Dest}(M,t)}{dt}\right)dM + \left( \frac{dN_{Form}(M,t)}{dt}\right) dM
\label{eqn17}
\end{align}

We have already given the corrections for the formation rate $(dN_{Form}/dt)dM$
and the destruction rate $(dN_{Dest}/dt)dM$ in the last two sections. The 
correction in the rate of change of number density $(dN_{PS}(M,t)/dt)dM$ can 
be written in terms of the correction in the formation and destruction rates.
In Figure (7) and  Figure (8) we show the rate of change of the comoving number 
density and the exact and approximate correction terms in it for the power law 
and LCDM  model  respectively. From equation~(\ref{eqn17}) it is clear that for 
$\sigma(M)<\delta_c/D(t)$  the rate of change of number density ${\dot N}_{PS}$  
is dominated by the formation rate, and for  $\sigma(M) > \delta_c/D(t)$ by the 
destruction  rate. For any time  $t$, we can find a mass scale $M_c$ for which
 $\sigma(M_c)= \delta_c/D(t)$ i.e., the formation  and the destruction  rate are 
equal and so there is no net change in the comoving  number density of objects at that 
scale. In hierarchical clustering models i.e., $\sigma(M)$ is a decreasing 
function of mass, comoving number density at large scales mainly changes due 
to the formation of massive haloes and at small scales due to destruction of 
smaller haloes.

The rate of change  in the number density is underestimated at large and small
scales, however, it is overestimated at intermediate scales. This feature is 
clear from Figure (7) in which the actual and approximate error terms are 
positive at large and small scales but they are negative at intermediate
scales and we have two zeroes crossing.

\vspace{1.0cm}

{\centering{\section{Discussion}}}

In the hierarchical clustering models of structure formation  formation and 
destruction of haloes is a common  process. In the present study we have shown
that the formation and destruction rate of haloes due to gravitational 
clustering are affected significantly if the size of the simulation box is not 
sufficiently large. On the basis of the Press-Schechter formalism we have 
given the analytic expressions for the corrections in the comoving number 
density, formation rate, destruction rate and the rate of change in the number
density of haloes at  a given mass scale. We have considered the implications 
of our analysis for the power law ($n=-2$) and LCDM models. Since the box 
corrections are more important for models which have significant power at 
large scales, so  most of the models in which there is none  or very less power
at large scales are not affected by the size of the simulation box. However,
models in which there is a lot of power at large scales  ($n$ is large and  
negative) the box effects can be quite large. In both the cases i.e., power law
and LCDM models, the scales at which we have shown the corrections are far 
below the size of the simulation box.

\vspace{0.5cm}

The main conclusions of the present study are as follows:

\vspace{0.5cm}

\begin{itemize}
\item
If the size of the simulation box in N-body simulations  is not large enough
then the clustering amplitude is underestimated at all scales.
\item
At any given time, there is a scale above which merging is dominated by the 
formation rate and below which it is dominated by the destruction rate. 
\item
The comoving number density of haloes is underestimated at large scales and 
overestimated at small scales when we reduce the size of the simulation box.
\item 
The formation and destruction rate also get modified  by reducing the size
of the simulation box. Particularly, they are underestimated at large scales 
and overestimated at small scales.
\item
The suppression of the formation rate as well as destruction rate at large 
scales is mainly due to the absence of fluctuations in the initial density 
field at large scales due to limitation of box size.
\item
In N-body simulations which have small power at large scales the corrections 
due to box size can be ignored. 
\end{itemize}

Observations suggest that cosmological perturbations were present at all
scales in the initial density field that have been probed.  Particularly in 
dark matter models  (Diemand {\it\ et al} 2005; Diemand {\it\ et al} 2006) 
the index of the power spectrum at small scales becomes close to $-3$. So in 
order to simulate these models one has to be very careful in choosing the size
of the simulation box. This is because if the index of power spectrum is close
to $-3$ at the  box scale also or if the scale of non-linearity is close to 
the size of the simulation box then the results can be significantly affected 
by the finite box effects.

In this and  Bagla \& Prasad (2006) we have presented our analytic results 
for taking into account the effects of finite box size on the mass function,
formation and destruction rate and other physical quantities. In the next 
paper of this series we will present a detailed  comparison of our analytic 
results with a set of cosmological N-Body simulations of different box size.

{\centering{\section{Acknowledgment}}}

I would like to thank Jasjeet Bagla for insightful comments and discussion. 
Numerical work for this study was carried out at cluster computing facilities 
at the Harish-Chandra Research Institute (http://www.cluster.mri.ernet.in). 
This research has made use of NASA's Astrophysical  Data System.

\label{lastpage}
\end{document}